\documentclass{ifacconf}
\usepackage{enumerate}
\usepackage{graphicx}      
\usepackage{natbib}        
\usepackage{amsmath}
\usepackage{amsfonts}
\usepackage{amssymb}
\usepackage{bbm}
\usepackage{xcolor}
\usepackage{tikz}
\usepackage{pict2e}
\usepackage{soul}
\usepackage{dsfont}
\tikzstyle{v_c}=[circle, draw,inner sep=2pt, minimum width=12pt]
\tikzstyle{v_a}=[circle, draw,inner sep=2pt, minimum width=12pt, color=red]
\tikzstyle{edge} = [draw,thick,-,font=\small ]
\tikzstyle{label} = [draw,fill=black,font=\normalsize]

\newcommand{\R}{\mathbb{R}}

\newcommand{\B}{\mathcal{B}}

\newcommand{\X}{\mathcal{X}}

\newcommand{\V}{\mathcal{V}}
\newcommand{\A}{\mathcal{A}}

\newcommand{\ut}{\set{u_i}_{i\in \mathcal{V}}}

\newcommand{\abs}[1]{\left\vert#1\right\vert}
\newcommand{\set}[1]{\left\{#1\right\}}

\newcommand{\argmax}{\operatorname{argmax}}

\newtheorem{theorem}{Theorem}

\newtheorem{proposition}[theorem]{Proposition}

\begin{document}
\begin{frontmatter}

\title{On games with coordinating and anti-coordinating agents} 

	\thanks[footnoteinfo]{Giacomo Como is also with the Department of Automatic Control, Lund University, Sweden. This work was partially supported by MIUR grant Dipartimenti di Eccellenza 2018--2022 [CUP: E11G18000350001], the Swedish Research Council, and by the Compagnia di San Paolo.}

\author[First]{Martina Vanelli} 
\author[First]{Laura Arditti} 
\author[First]{Giacomo Como}
\author[First]{Fabio Fagnani}

	\address[First]{Department of Mathematical Sciences ``G.L.~Lagrange'', Politecnico di Torino, Corso Duca degli Abruzzi 24, 10129 Torino, Italy\\ 
 (e-mail: \{martina.vanelli,laura.arditti,giacomo.como,fabio.fagnani\}@polito.it).}


\begin{abstract}                
This work studies Nash equilibria for games where a mixture of
coordinating and anti-coordinating agents, with possibly heterogeneous thresholds, coexist and interact through an all-to-all network. Whilst games with only coordinating or only anti-coordinating agents are potential, also in the presence of heterogeneities, this does not hold when both type of agents are simultaneously present. This makes their analysis more difficult and existence of Nash equilibria not guaranteed.
Our main result is a checkable condition on the threshold distributions that characterizes the existence of Nash equilibria in such mixed games. When this condition is satisfied an explicit algorithm allows to determine the complete set of such equilibria. Moreover, for the special case when only one type of agents is present (either coordinating or anti-coordinating), our results allow an explicit computation of the cardinality of Nash equilibria.


\end{abstract}


\end{frontmatter}

\section{Introduction}
 
In games of strategic complements, the best response action of a player is increasing in the action of the other players. Examples of such games include the adoption of a new technology, beliefs or behavioral attitudes in social influence systems, or cooperative interchanges among economical actors. In games of strategic substitutes just the opposite happens, the best response action of a player is decreasing in the action of the other players. Applications of such models include, for example, local public good provision, information gathering, firms interacting in competitive markets.

While these two classes of games have received considerable attention in recent literature  (\cite{gamesonnetworks}, \cite{bramoulle}), relatively unexplored are mixed models where these two strategic interactions coexist.  However, examples of social or economic model  where such behaviors coexist are rather frequent, e.g., collecting actions, interacting firms with cooperative and competitive features. 

In this paper, we focus on a particular instance of such heterogeneous models. We consider games  where actions are all binary $\{-1, +1\}$ and players are split into two classes: one of coordinating agents and one of anti-coordinating agents. That of coordination and anti-coordination games are two basic examples of games with respectively strategic complements and strategic substitutes. Coordinating agents have an incentive for action $1$ when such action is played by a fraction of the remaining players above a certain threshold while anti-coordinating agents have an incentive for action $1$ when such action is played by a fraction of the remaining players below a certain threshold. We will refer to such models as mixed coordination/anti-coordination (CAC) games.

Our aim is to study Nash equilibria in mixed CAC games. Pure coordination games and pure anti-coordination games always admit Nash equilibria as they are potential games (\cite{monderer}) also when the thresholds are heterogeneous. Instead general mixed CAC games are no longer potential and the existence of Nash equilibria is not guaranteed. Furthermore, even when existence is guaranteed, the set of  Nash equilibria is in general unknown. For the special case of coordinating agents with identical thresholds, a characterization of the Nash equilibria was proposed by \cite{morris}. 

There is a related literature where the best-response dynamics is analyzed, in particular its convergence to Nash equilibria. \cite{cao} proved that the asynchronous best-response dynamics of pure coordination games and pure anti-coordination games will almost surely converge to Nash equilibria for every network topology and every set of thresholds. \cite{granovetter} studied the synchronous best-response dynamics of coordinating agents as a linear threshold model. Similar results have been generalized to configuration models by \cite{thresholdcascades} and to games with a mixture of coordinating and anti-coordinating agents by \cite{fen}. Other results concern time of convergence of asynchronous best-response dynamics and its dependence on the network structure (\cite{ellison}, \cite{kandori}, \cite{montanari}).

Our contribution is a throughout analysis on the existence and the structure of pure strategy Nash equilibria for mixed coordination/anti-coordination (CAC) games with possibly heterogeneous activation thresholds. 
In particular we give the following contributions: (i) We establish a checkable necessary and sufficient condition involving the cumulative distribution functions of the thresholds for the existence of NE of a general mixed CAC game; and 
(ii) In the special case when only coordinating or anti-coordinating agents are present we classify all NE and determine their cardinality. While existence was well known in the literature, to the best of our knowledge such precise characterization was not yet presented.

The rest of the paper is organized as follows. Section \ref{model} is devoted to the description of the model and Section \ref{main_result} presents the main results. We conclude with Section \ref{conclusion} containing a summary and further work. 

\section{Model description} \label{model}
In this section, we formally define the mixed CAC game, we explicit the best response function and we make some general observations on the existence of Nash equilibria. 
In order to give an intuition of the upcoming results, in the end we briefly study the case of an infinite population. 

\subsection{Definition of the game}
Consider a finite set of agents $\V = \{1,\dots,n\}$ partitioned into two classes:
$\V_c\cup \V_a= \V$, $\V_c \cap \V_a = \emptyset$. Each agent $i\in \V$ is described by two elements: the type $\delta_i$ describing the class it belongs to
$$
\delta_i := \begin{cases}
1 \hspace{1cm} &\text{if }i\in \V_c \\
-1  &\text{if }i\in \V_a
\end{cases}
$$
and its personal weight $d_i \in \R$.

The action set of all agents is the binary set $\A:= \set{-1,+1}$. We denote with $x_i\in\A$ the action chosen by agent $i$ and with $x\in \X:= \A^\V$ the vector assembling the actions of all players (action profile). As usual in game theory, $x_{-i}\in \A^{\V\setminus\set{i}}$ is the vector obtained from the action profile $x\in \X:= \A^\V$ by removing its $i^{\text{th}}$ entry. The utility of agent $i$, when she plays action $x_i$ and the remaining players $x_{-i}$, is given by
\begin{equation}\label{eq:utilities_mixed_game}
	u_i(x_i, x_{-i}) :=
	\delta_i\left(\sum_{j\neq i} x_ix_j -d_i x_i\right)\,,\hspace{0.3cm}i\in \V
\end{equation} 


 

Agents in $\V_c$ ($\delta_i=1$) are  \textit{coordinating agents} (their utility increases with the number of agents playing their same action), while agents in $\V_a$ ($\delta_i=-1$) are  \textit{anti-coordinating agents} (their utility decreases with the number of agents playing their same action).
%
%
%

The weight $d_i$ represents the tendency of agent $i$ in choosing an action over the other. Indeed, the sign of weight $d_i$, along with  $\delta_i$, determines which is the risk dominant action, i.e., which is the best action for agent $i$ in the absence of any external influences. If $d_i =0$ the two actions are called \textit{risk-neutral}.

\textit{Definition.} The \textit{mixed coordination/anti-coordination} \\(CAC) game is a game with agent set $\V$, action configuration space $\X$ and utilities
$\ut:\X \rightarrow \R$ in (\ref{eq:utilities_mixed_game}).

In the special cases when only one type of agents is present, namely $\V_c = \V$ or $\V_a=\V$, we will simply call it, \textit{coordination game} or, respectively, \textit{anti-coordination game}.


\subsection{Best response and Nash equilibria}

In strategic games, agents are assumed to be rational, i.e., they choose their action with the aim of maximizing the utility. Given the actions of the others, the \textit{best response} (BR) function returns the set of the best actions for agent $i$, that is, the actions that achieve the highest utility. 
Formally, the BR is given by 
$$\B_i(x_{-i})= \argmax_{x_i \in \A}u_i(x_i,x_{-i}) \,.$$
For the purpose of our study, it is convenient to express the best response function in terms of the number of opponents playing action $+1$, that is, $$n_i^{+}(x)=\abs{\set{j\in\V\setminus\set{i}\mid x_j=+1}}\,.$$
From (\ref{eq:utilities_mixed_game}) and the fact that $\sum_{j\neq i} x_j =2n^+_i(x)-(n-1)$, it follows that the BR of an agent $i\in \V$ can be written in the following way:
	\begin{equation}\label{eq:br}
	\B_i(x_{-i}) =
	\begin{cases} \set{+1}  \hspace{0.3cm}
	\text{if }\delta_i(n_i^+(x)-r_i(n-1)) >0 \\
	\set{-1} \hspace{0.3cm}\text{if }\delta_i(n_i^+(x)-r_i(n-1))<0 \\
	\set{\pm 1} \hspace{0.3cm}\text{if }\delta_i(n_i^+(x)-r_i(n-1)) = 0
	\end{cases} 
	\end{equation}
where $r_i: = \frac{1}{2} + \frac{d_i}{2(n-1)}$.


In words, when $\delta_i = 1$, the best action is $+1$ if more than $r_i(n-i)$ opponents are playing $+1$, whereas, when $\delta_i=-1$, the best action is $+1$ if less than $r_i(n-i)$ opponents are playing $+1$. 
Hence, the best response changes depending if the fraction of opponents playing action $+1$, that is, $n_i^+(x)/(n-1)$, is above or below the \textit{threshold} $r_i$. 
Agents having $r_i>1$ or $r_i<-1$ are called \textit{stubborn agents}, as their best response function is constantly equal to either $\set{-1}$ or $\set{+1}$, independently of the actions of others.


A \textit{Nash equilibria} is a configuration $x \in \X$ such that $x_i\in \B_i(x_{-i})$ for any agent $i\in\V$. In the following we will denote by $\mathcal{N}$ the set of Nash equilibria of the game.

Both the coordination game and the anti-coordination game, for any possible choice of the weights, are potential games (\cite{monderer}). This fact is very well known in case of homogeneous thresholds, but actually holds in general. Indeed, if we define the two functions
	\begin{equation}\label{potential_function_het_coord}
	\Phi_c(x)=\frac{1}{2}\sum_{\overset{i,j\in \V}{i\neq j}}x_i x_j - \sum_{i\in \V} d_i x_i 
	\end{equation}
	\begin{equation}\label{potential_function_het_anti_coord}
	\Phi_a(x)=-\Phi_c(x)
	\end{equation} 
we can derive from (\ref{eq:utilities_mixed_game}) that, depending if the game is a coordination or an anti-coordination one, it respectively holds
\begin{equation}\label{potential_function}
	\begin{array}{rcl}u_i(y_i, x_{-i})-u_i(x_i, x_{-i}) &=& \Phi_c(y_i, x_{-i})-\Phi_c(x_i, x_{-i})\\
	u_i(y_i, x_{-i})-u_i(x_i, x_{-i}) &=& \Phi_a(y_i, x_{-i})-\Phi_a(x_i, x_{-i})\end{array}
\end{equation}
for all $x_i,y_i \in \A$, $x_{-i}\in \A^{\V\setminus \set{i}}$. 

The potential property guarantees the existence of at least one Nash equilibrium, which is a maximum point of the potential function. If there are no stubborn agents, the two consensus configurations, namely $\mathds{1}$ and $-\mathds{1}$, are NE of the coordination game for any set of weights. This is a peculiar property of coordination games and, in general, an explicit characterization of all the NE is hard to find.

It is enough for a coordinating agent to interact with an anti-coordinating agent to lose the potential property. 
\begin{prop}\label{pot_mixed}
	If $\V_c \neq \emptyset$ and $\V_a \neq \emptyset$ the mixed CAC game is \textit{not} a potential game.
\end{prop}


Hence, the study of Nash equilibria for the mixed CAC game is a challenging problem and not even the existence of such configuration is guaranteed, as the following simple example shows.
\subsubsection{Example 1.}
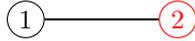
\begin{figure}[h]
	\centering
	\begin{tikzpicture}[scale=1, swap]
	\node[v_c] (vc) at (0,0) {1};
	\node[v_a] (va) at (2,0) {2};
	\draw[edge] (vc) to (va);
	\end{tikzpicture}
	\caption{The discoordination game. }
	\label{fig:disc_game}
\end{figure}

Let us consider the very simple case of a two-player game with $\V_c=\set{1}$ and $\V_a=\set{2}$ where both agents have weights $d_i = 0$ (Fig. \ref{fig:disc_game}). This is known as the \textit{discoordination game} or \textit{matching pennies} game, since it describes the situation where one agent aims at coordinating while the other wants to play the opposite action to that of her opponent.
The discoordination game admits no Nash equilibria. 

\subsection{The formalism of distribution functions}


We now set some notation that will be used throughout the rest of the paper.

\begin{itemize}
\item  $n_c =|\V_c|$ and $n_a=|\V_a|$ are the number of coordinating and anti-coordinating agents respectively, while $\alpha=n_c/n$ is the fraction of coordinating agents.
\item Given a configuration $x\in\X$, we denote by $z(x)$, $z_c(x)$, $z_a(x)$ the fraction of agents that in $x$ are playing action $1$ in, respectively, the total population, in the set of coordinating agents, and in the set of anti-coordinating agents:
%
%
%
%
\begin{equation}\label{z}
\begin{array}{rcl}z(x) &:=& n^{-1}|\set{i\in \V \mid x_i=+1}|,\\
z_c(x) &:=& n_c^{-1}|\set{i\in \V_c \mid x_i=+1}|, \\
z_a(x) &:=& n_a^{-1}|\set{i\in \V_a \mid x_i=+1}|\end{array}
\end{equation}
It holds
\begin{equation*}
z(x) = \alpha z_c(x) + (1-\alpha) z_a(x)
\end{equation*}
\end{itemize}

Furthermore, let
\begin{equation}\label{F}
F(z) := \frac{1}{n}\abs{\set{i \in \V \mid r_i \leq z}}
\end{equation} 
be the \textit{threshold cumulative distribution function} (CDF) that returns the fraction of agents having threshold \textit{less} than or \textit{equal} to $z$. 

In general, the threshold CDF is non-decreasing, piece-wise constant and continuous to the right with discontinuities occurring at points $r_i$, $i\in \V$. We denote by $$F^-(z):=\lim_{a \rightarrow z^{-}} F(a) =\frac{1}{n}\abs{\set{i \in \V \mid r_i < z}} $$
 the left limit of $F$ in $z$.

Similarly, the \textit{threshold complementary cumulative distribution function}(CCDF) is given by \begin{equation}\label{F}
G(z) := \frac{1}{n}\abs{\set{i \in \V \mid r_i > z}}
\end{equation} and returns the fraction of agents having threshold \textit{greater} than a given value.
Note that $G(z) =  1- F(z)$. Accordingly, the threshold CCDF is non-increasing, piece-wise constant and continuous to the right with discontinuities occurring at points $r_i$, $i\in \V$.  We denote by $G^-(z):=\lim_{a \rightarrow z^{-}} G(a)$
the left limit of $G$ in $z$.


Throughout the paper, we denote the threshold CDF of the coordinating agents by $F_c$ and  threshold CCDF of the anti-coordinating agents by $G_a$. Our goal is to characterize Nash equilibria in terms of the thresholds CDF and CCDF. Part of the difficulty is due to the fact that in our finite size setting, the fraction variables $z$, $z_c$, and $z_a$ are not continuous and variations restricted to multiples of $1/n$. To give an intuition of the results that will be presented in the next section, below we briefly analyze the case of an infinite population of agents where such technical issues disappear.

\subsection{Infinite population}

We assume an infinite continuous population of agents with a percentage $\alpha\in [0,1]$ of coordinating agents, while the remaining agents are anti-coordinating.
Furthermore, we assume that $F_c:\R \rightarrow [0,1]$ and $G_a:\R \rightarrow [0,1]$ are \textit{continuous} functions.

A configuration corresponding to a global fraction of $1$ given by $z^*$  is a Nash if and only if the players in $\V_c$ playing action $1$ are exactly those having threshold above $z^*$ and all players in $\V_a$ playing action $1$ are exactly those having threshold below $z^*$. Indicating with $z^*_c\,,z^*_a$ the fraction of agents playing action $1$ in respectively $\V_c$ and $\V_a$, Nash equilibria can then be characterized by the equations
\begin{equation}\label{system:continuous}
\begin{cases}
z^*_c  = F_c(z^*) \\
z_a^*  = G_a(z^*) \\
z^* = \alpha z_c^* +(1-\alpha)z_a^*
\end{cases}
\end{equation}
If we define
	\begin{equation}\label{H_alpha}
	H_\alpha(z):=\alpha  F_c(z)  +(1-\alpha)G_a(z)\,,
	\end{equation}
	we can see that all the solutions of (\ref{system:continuous}) can be found by first solving
	\begin{equation}\label{eq:cond_continuous}
	z^*= H_{\alpha}(z^*)
	\end{equation}
	and then considering $z^*_c = F_c(z^*)$ and $z^*_a = G_a(z^*)$.
	Notice that since $H_\alpha:[0,1] \rightarrow [0,1]$ is continuous, solutions to (\ref{eq:cond_continuous}) always exist.
%
%
%
%
This shows that the mixed CAC game for an infinite continuous population and assuming that $F_c$ and $G_a$ are continuous, always admits Nash equilibria.


\begin{figure}
	\centering	\includegraphics[width=0.47\linewidth]{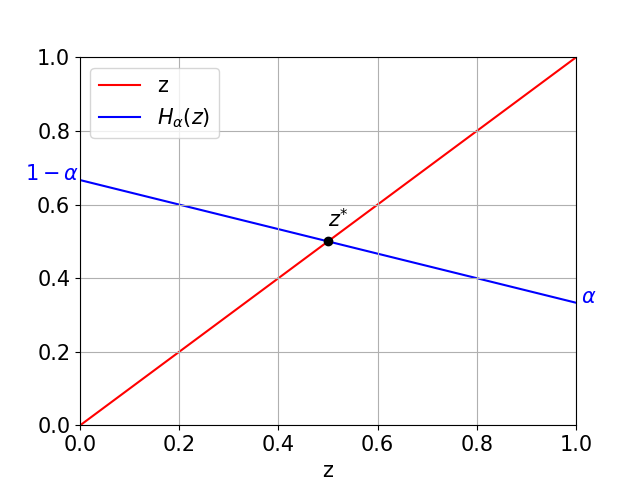}\hspace{0.03\linewidth}	\includegraphics[width=0.47\linewidth]{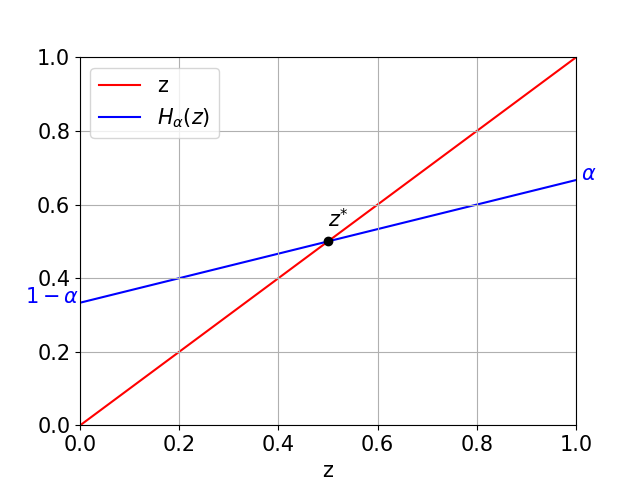}
	\caption{We study the fixed point equation (\ref{eq:cond_continuous}) for Ex. 2. On the left we consider $\alpha<\frac{1}{2}$, on the right $\alpha>\frac{1}{2}$.}
	\label{fig:cont1}
\end{figure}

\subsubsection*{Example 2.}
Let the thresholds of the coordinating agents be \textit{uniformly distributed} in $[0,1]$, as well as the thresholds of the anti-coordinating agents, i.e. $F_c(z) = z\mathds{1}_{[0,1]}(z)+\mathds{1}_{(1,\infty)}(z)$ and $G_a(z) = 1-F_c(z)$. Then $$H_\alpha(z) = (2\alpha-1)z + (1-\alpha)$$
and its unique fixed point is $z^* = \frac{1}{2}$ (Fig. \ref{fig:cont1}). The NE is given by $z^*_c = F_c(\frac{1}{2}) = \frac{1}{2}$ and $z^*_a = G_a(\frac{1}{2}) = \frac{1}{2}$.


\section{Nash equilibria of mixed CAC games}\label{main_result}
In this section, we investigate the existence, the uniqueness and the characterization of NE of mixed CAC games. 

Our main result is to provide a \textit{necessary} and \textit{sufficient} condition for an action configuration to be a Nash equilibrium of the mixed CAC game. The condition, which is a slight modification of (\ref{system:continuous}), only involves the fractions of agents playing action $+1$ in the whole population and in the subsets $\V_c$ and $\V_a$. 

As we shall see, the condition is checkable and it reduces the complexity of the search of all NE to analyze $n+1$ triples of candidate fractions.
%
%
%
%
%
%
%
%
%
In the special case where only coordinating or anti-coordinating agents are present, we will be able to give more explicit results.
%

\subsection{Formal statement of the main result}

In this section, we formally state the main result.
Given a Nash equilibrium $x^*\in \X$, we shortly denote $z^*:= z(x^*)$, $z^*_c:= z_c(x^*)$ and $z^*_a:= z_a(x^*)$.

\begin{theorem}\label{prop:nec_suff_cond}
	Consider the mixed CAC game.
	\begin{enumerate}[(i)]
		\item 	A Nash equilibrium $x^*\in \X$ satisfies 
		\begin{equation}\label{eq:nec_suff_cond}
		\hspace{-0.54cm}\begin{cases}
		F^-_c\left(\frac{n}{n-1}z^*\right) =z^*_c=F_c\left(\frac{n}{n-1}(z^*-\frac{1}{n})\right)\\
		G^-_a\left(\frac{n}{n-1}(z^*-\frac{1}{n})\right) \geq z_a^*  \geq G_a\left(\frac{n}{n-1}z^*\right) \\
		z^* = \alpha z_c^* +(1-\alpha)z_a^*
		\end{cases}
		\end{equation}
		\item Given $z^* \in \set{0,\frac{1}{n},\dots, 1}$, $z^*_c\in \{0,\frac{1}{n_c},\dots, 1\}$, $z^*_a \in \{0,\frac{1}{n_a},\dots, 1\}$ satisfying $(\ref{eq:nec_suff_cond})$, there exists at least one Nash equilibrium $x^*\in \X$ such that $z(x^*)=z^*$, $z_c(x^*)=z_c^*$ and $z_a(x^*)=z_a^*$. 
	\end{enumerate}


\end{theorem} 
\subsubsection*{Remark 1.} 
The Nash equilibrium whose existence is guaranteed by (ii) can be built in the following way. First, we set the actions of the coordinating agents $i\in \V_c$ accordingly to the following formula
\begin{equation}\label{eq:build_eq_coord}
x^*_i = \begin{cases}
+1 \quad &\text{if } r_i \leq \frac{n}{n-1}\left(z^* - 
\frac{1}{n}\right) \\
-1  &\text{if } r_i \geq \frac{n}{n-1}\,z^*
\end{cases}
\end{equation}
Then, we force the actions of the anti-coordinating agents $i\in \V_a$ to satisfy
\begin{equation}\label{eq:build_eq_anti_coord}
x^*_i = \begin{cases}
-1 &\text{if } r_i < \frac{n}{n-1}\left(z^* - 
\frac{1}{n}\right) \\
+1 &\text{if } 
 r_i > \frac{n}{n-1}\,z^* \,.
\end{cases}
\end{equation}
Finally, we set the actions of the anti-coordinating agents $i\in \V_a$ having thresholds $r_i \in [\frac{n}{n-1}(z^* - 
\frac{1}{n}), \frac{n}{n-1}\,z^* ]$ in such a way that the condition $z^*(x^*)=z^*$ is met. In general, the associated Nash equilibrium is not unique.

Theorem \ref{prop:nec_suff_cond} gives a tool to investigate the existence of Nash equilibria of the mixed CAC game and it permits to characterize all the NE of the game. Indeed, 
once we find all triples of fractions $(z^*,z^*_c,z^*_a)$ satisfying $(\ref{eq:nec_suff_cond})$, then we can trivially find all the Nash equilibria of the game.

\subsubsection{Remark 2.} Consider a triple $(z^*,z^*_c,z^*_a)$ satisfying (\ref{eq:nec_suff_cond}). The values of $z^*_c$ and $z^*_a$ are uniquely determined from $z^*$ through the $1^{\text{st}}$ and $3^{\text{rd}}$ equations. Hence, there are at most $n+1$ triples satisfying the conditions in (\ref{eq:nec_suff_cond}).
 

\subsection{Nash equilibria of the coordination game}
In this section, we focus on the coordination game, i.e., $\V_c = \V$. 
A straightforward substitution of $\alpha=1$ in (\ref{eq:nec_suff_cond}) leads to the following characterization of NE for coordination games.
\begin{cor}\label{prop:cond_coord}
	Consider the coordination game. 
	\begin{enumerate}[(i)]
		\item A Nash equilibrium $x^*\in\X$ satisfies 	
		\begin{equation}\label{eq:cond_coord}
		 F^-_c\left(\frac{n}{n-1}z^*\right) =z^*  =F_c\left(\frac{n}{n-1}(z^*-\frac{1}{n})\right)
		\end{equation}
		\item Given $z^*\in\set{0,\frac{1}{n},\dots, 1}$ satisfying (\ref{eq:cond_coord}), there exists a Nash equilibrium $x^*\in\X$ such that $z(x^*)=z^*$.
	\end{enumerate}
\end{cor}

In this case, we can strengthen the analysis as reported in the result below.

\begin{proposition}\label{pr:ex_card_coord} The coordination game always admits Nash equilibria. Moreover, the number of NE coincide with the number of solutions $z^*$ of the equations (\ref{eq:cond_coord}) and this number never exceeds $n+1$.
\end{proposition}
We comment on this result. Existence follows from the fact that the game is in this case potential, but can also directly derived from condition (\ref{eq:cond_coord}). 
Notice that, if there are no stubborn agents, that is, $F(0)=0$ and $F(1)=1$, then $z^0= 0$ and $z^1=1$ are solutions of (\ref{eq:cond_coord}). They correspond to the two \textit{consensus configurations}.
On the other hand, the fact that a NE $x^*$ is completely determined by the corresponding value of $z^*$ is an immediate consequence of the characterization expressed in formula (\ref{eq:build_eq_coord}).

The following example shows a case where equations (\ref{eq:cond_coord}) has exactly  $n+1$ distinct solutions.
\subsubsection*{Example 3.}
\begin{figure}
	\centering
	\includegraphics[width=0.49\linewidth]{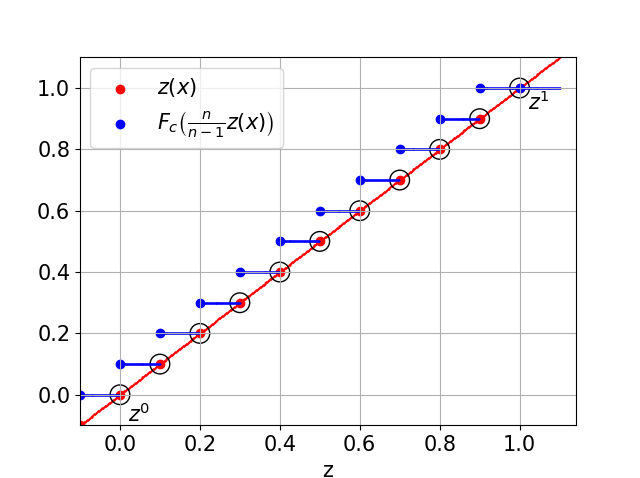}
	\includegraphics[width=0.49\linewidth]{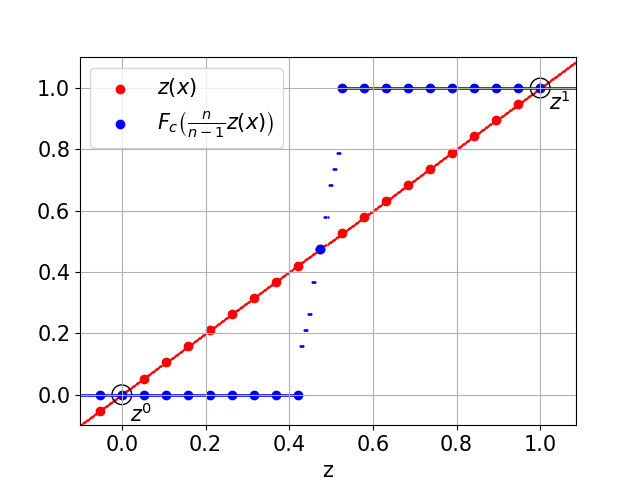}
	
	\caption{On the left, for thresholds in Ex. 3, (\ref{eq:cond_coord}) admits $n+1$ solutions. On the right, we consider an example where $z^0=0$ and $z^1=1$ are the unique solutions of (\ref{eq:cond_coord}). Even if $z^*=\frac{n-1}{2n}$ is a fixed point, the CDF is not flat in the interval $[z^*-\frac{1}{n}, z^*)$. }
	\label{fig:eq_coord}
\end{figure}

Consider the thresholds $r = [ 0,\frac{1}{n-1},\dots, 1]$. The condition in (\ref{eq:cond_coord}) admits the $n+1$ solutions $0,\frac{1}{n},\dots,1$ as shown in Fig. \ref{fig:eq_coord}.


Finally, notice that a fixed point $z^* = F^-_c(\frac{n}{n-1}z^*)$ is a solution of (\ref{eq:cond_coord}) only when there is a flat interval preceding the fixed point large at least $\frac{1}{n}$. In Fig. 3, we provide an example where a fixed point does not satisfy (\ref{eq:cond_coord}).

\subsection{Nash equilibria of the anti-coordination game }
In this section, we focus on the anti-coordination game, i.e., $\V_a = \V$. 
A straightforward substitution of $\alpha=0$ in (\ref{eq:nec_suff_cond}) leads to the following characterization of NE for anti-coordination games.
\begin{cor}\label{prop:cond_anti_coord}
	Consider the anti-coordination game. 
	\begin{enumerate}[(i)]
		\item A Nash equilibrium $x^*\in\X$ satisfies 
		\begin{equation}\label{eq:cond_anti_coord}
			G^-_a\left(\frac{n}{n-1}(z^*-\frac{1}{n})\right) \geq z^*  \geq G_a\left(\frac{n}{n-1}z^*\right)
		\end{equation}
		\item Given $z^*\in\set{0,\frac{1}{n},\dots,1}$ satisfying (\ref{eq:cond_anti_coord}), there exists a Nash equilibrium $x^*\in\X$ such that $z(x^*)=z^*$. 
	\end{enumerate}

\end{cor}

A deeper insight is possible also for this case, though it has peculiar differences with respect to the pure coordination case. We first note that a \textit{fixed point} $z^*\in \set{0,\frac{1}{n},\dots, 1}$ of the threshold CCDF, namely $z^*= G_a(\frac{n}{n-1}z^*)$, satisfies (\ref{eq:cond_anti_coord}). Such fixed points however not necessarily exist and, in any case, other solutions to (\ref{eq:cond_anti_coord}) may show up. 
In the following result, we propose a sufficient existence condition for NE that encompasses the fixed point condition above.

\begin{figure}
	\centering
	\includegraphics[width=0.49\linewidth]{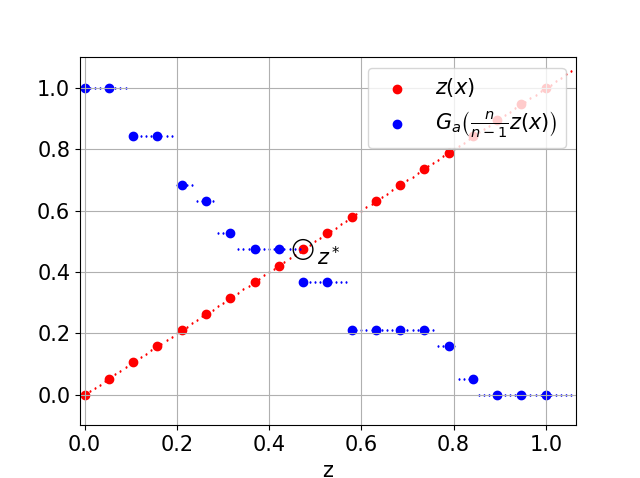}
	\includegraphics[width=0.49\linewidth]{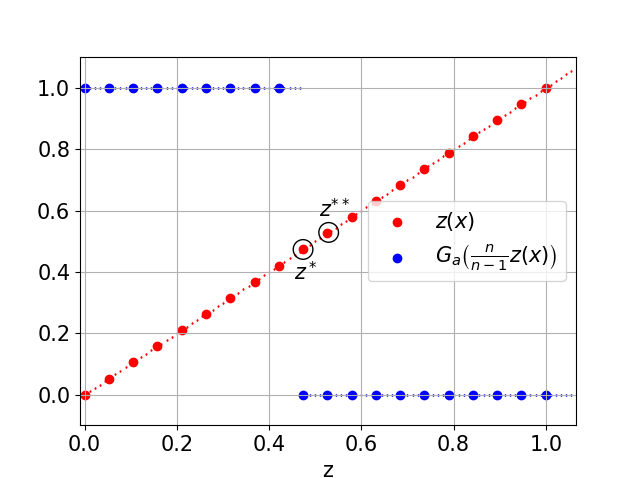}
	\caption{On the left, an example when inequality (\ref{eq:cond_anti_coord}) admits the unique solution $z^* =\frac{9}{18}$. On the right, we consider $n=19$ and $r_i =\frac{1}{2}$, for all $i\in \V$. The inequality (\ref{eq:cond_anti_coord}) admits the two solutions $z^*=\frac{9}{19}$ and $z^{**}=\frac{10}{19}$.}
	\label{fig:eq_anti_coord2}
\end{figure}

\begin{proposition}\label{pr:ex_z_anticoord} Consider the inequalities
\begin{equation}\label{eq:z_cond_anti_coord}
\begin{cases}z^*-\frac{1}{n} <G_a\left(\frac{n}{n-1}(z^*- \frac{1}{n})\right)\\ z^* \geq G_a\left(\frac{n}{n-1}z^*\right)\,.\end{cases}
\end{equation}
Then, any $z^*\in \set{0,\frac{1}{n},\dots, 1}$ satisfying (\ref{eq:z_cond_anti_coord})  is a solution of (\ref{eq:cond_anti_coord}).
\end{proposition}
Since $G_a$ is non-increasing with codomain $[0,1]$, solutions to (\ref{eq:z_cond_anti_coord}) always exist. An example is shown in Fig. \ref{fig:eq_anti_coord2}. This argument in particular proves the existence of NE for anti-coordination game, fact that could also be derived from the fact that such games are potential. Notice that, concretely, starting from $z^*$ a corresponding NE $x^*$ can be constructed setting the actions of part of the agents as in (\ref{eq:build_eq_anti_coord}) and the remaining actions in such a way that the condition $z^*(x^*)=z^*$ is met. The associated Nash equilibrium is, in general, \textit{not} unique.

The next result furtherly refines our analysis of the NE of the anti-coordination game and includes an analytical computation of its cardinality.

\begin{proposition}\label{prop:card_anticoord}For the anti-coordination game, the following facts hold true.
\begin{enumerate}
\item The number of solutions of the conditions (\ref{eq:cond_anti_coord}) is either $1$ or $2$.
\item Indicated with $\mathcal Z$ the set of solutions of (\ref{eq:cond_anti_coord}), the number of NE is given by
\begin{equation}
	\abs{\mathcal{N}}=\sum_{z \in \mathcal{Z} } \binom{n(\zeta_2-\zeta_1)}{n(z-\zeta_1)}
	\end{equation}
where
\begin{flalign}\label{z1_z2}
\begin{aligned}
\hspace{-0.3cm}\zeta_1 := G_a\left(\frac{n}{n-1}z\right), \zeta_2:= G^-_a\left(\frac{n}{n-1}(z-\frac{1}{n})\right).
\end{aligned}
\end{flalign}
\end{enumerate}
\end{proposition}

We briefly comment on the statement in item 1.: the possibility of a second solution in (\ref{eq:cond_anti_coord}) come from the fact that if $z^*$ is a solution and there is a sufficiently big jump between $G_a(z^*)$ and $G^-_a(z^*)$, then  $z^{**} =z^* + \frac{1}{n}$ is another solution. No other choice is possible as $G_a$ is by definition non-increasing. This phenomenon is shown in Fig. \ref{fig:eq_anti_coord2}.  

%

\subsection{Nash Equilibria of the mixed CAC game}
	In this section, we make some final considerations on the general mixed case. The difficulty in analyzing the existence condition (\ref{eq:nec_suff_cond}) is due to the simultaneous presence of the two threshold functions that give rise to two coupled conditions. In the infinite continuous population case this was essentially overcome by considering the function $H_\alpha$ defined in (\ref{H_alpha}) and reducing the problem to a fixed point investigation. Similar considerations can also be done in the finite size case and, even if they do not lead to a complete characterization, they do shed some light on the problem.
	To this aim, consider any $(z^*,z^*_c,z^*_a)$ satisfying (\ref{eq:nec_suff_cond}) and makes the extra assumption that
	\begin{flalign*}
	&F^-_c(\frac{n}{n-1}z^*) =F_c(\frac{n}{n-1}z^*)\,,	\\ &G^-_a\left(\frac{n}{n-1}(z^*-\frac{1}{n})\right)=G_a\left(\frac{n}{n-1}(z^*-\frac{1}{n})\right)\,,
	\end{flalign*} 
		Then, $z^*$ also satisfies the pair of inequalities
		\begin{equation}\label{eq:z_cond_coord_anti_coord}
	H_\alpha \left(\frac{n}{n-1}(z^*-\frac{1}{n})\right)\geq
	z^* \geq H_\alpha\left(\frac{n}{n-1}z^*\right)\,.
	\end{equation}
	This can be seen as an approximation of the fixed point relation $z^*=H_{\alpha}(z^*)$ we had in the infinite case. This argument says that, in determining the solutions of (\ref{eq:nec_suff_cond}),  all the possible candidates for the fraction $z^*$ are:
	\begin{enumerate}[(i)]
		\item the solutions of (\ref{eq:z_cond_coord_anti_coord}) that are continuous points of both $F_c(\frac{n}{n-1}\,\ast\,)$ and $G_a(\frac{n}{n-1}(\ast-\frac{1}{n}))$.
		\item The discontinuous points of either $F_c(\frac{n}{n-1}\,\ast\,)$ and $G_a(\frac{n}{n-1}(\ast-\frac{1}{n}))$.
		\end{enumerate}
	An example that shows how such argument can be concretely used is presented below.

\begin{figure}
	\centering
	\includegraphics[width=0.49\linewidth]{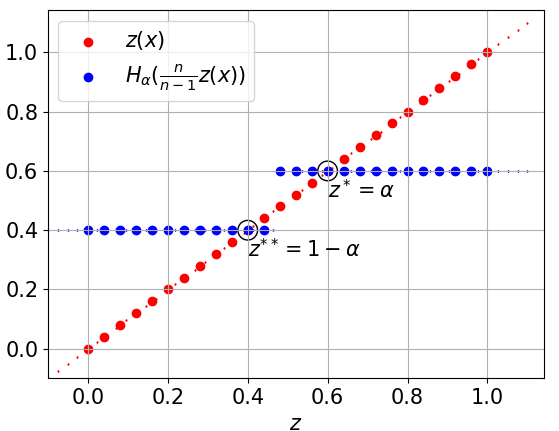}
	\caption{We consider Ex. 4 for $\alpha\geq \frac{1}{2}+\frac{1}{n}$. The pair of inequalities in (\ref{eq:z_cond_coord_anti_coord}) admits the two solutions $z^*=\alpha$ and $z^{**}=1-\alpha$.}
	\label{fig:eq_cohesive}
\end{figure}

\subsubsection*{Example 4 (Mixed majority/minority game).} Consider $r_i =\frac{1}{2}$ for all agents, and $\alpha \in(0,1)$. Then, we have
\begin{flalign*}
H_\alpha \left(\frac{n}{n-1}z\right)
= (1-\alpha)\mathds{1}_{[0, \frac{n-1}{2n})}(z) + \alpha\mathds{1}_{[\frac{n-1}{2n},1]}(z)\,. 
\end{flalign*}

Notice that $F_c(\frac{n}{n-1}\,\ast\,) $ is discontinuous in $\frac{n-1}{2n}$, whereas $G_a(\frac{n}{n-1}(\ast-\frac{1}{n}))$ is discontinuous in $\frac{n+1}{2n}$.
 
CASE 1: If $\alpha \geq \frac{1}{2}+ \frac{1}{n}$, the condition in $(\ref{eq:z_cond_coord_anti_coord})$ admits the two solutions $z^*=\alpha$ and $z^{**} = 1-\alpha$ (Fig. \ref{fig:eq_cohesive}). Hence, we have 4 candidates: $z^*$, $z^{**}$, and the 2 discontinuous points. 

	If we substitute the first candidate $z^* = \alpha$ in the $1^{\text{st}}$ and $3^{\text{rd}}$ equations of ($\ref{eq:nec_suff_cond}$), we get $z^*_c = 1 $ and $z^*_a= 0$. Note that the triple $(\alpha,1,0)$ does satisfy (\ref{eq:nec_suff_cond}). 
	The triple associated to $z^{**}$ is $(1-\alpha, 0,1)$ and it satisfies (\ref{eq:nec_suff_cond}) as well.
The two solutions correspond to the NE where all coordinating agents play action $1$, or action $-1$, respectively, while anti-coordinating agents play the opposite one. 

Notice that there are no other NE as the triples associated to $\frac{n-1}{2n}$ and $\frac{n+1}{2n}$ do not satisfy (\ref{eq:nec_suff_cond}).

\begin{figure}
	\centering
	\includegraphics[width=0.48\linewidth]{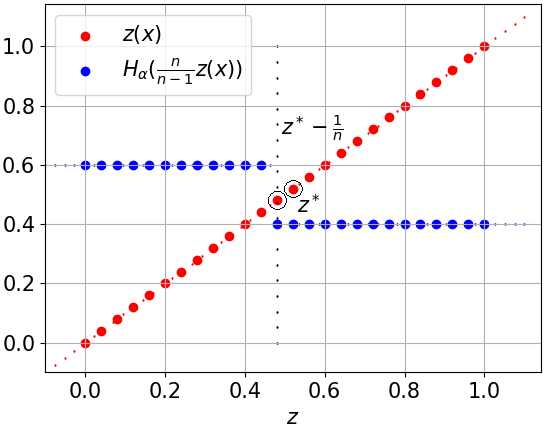}
	\includegraphics[width=0.49\linewidth]{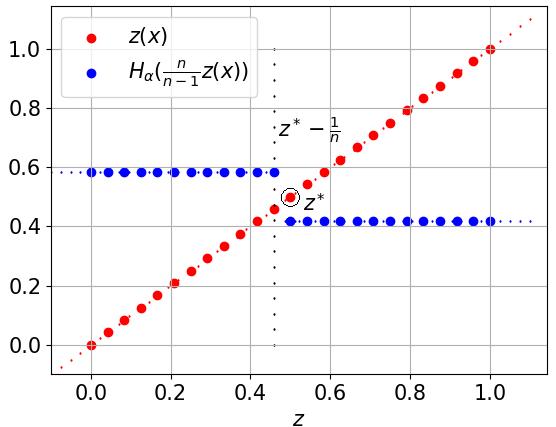}
	\caption{We consider Ex. 4 for $\alpha<\frac{1}{2}+\frac{1}{n}$. If $n$ is odd (left), the solutions of $(\ref{eq:z_cond_coord_anti_coord})$ are  $z^*=\frac{n-1}{2n}$ and $z^{**} =\frac{n+1}{2n}$. If $n$ is even (right), the unique solution is $z^*=\frac{1}{2}$. }
	\label{fig:eq_not_cohesive}
\end{figure}

CASE 2: If $\alpha < \frac{1}{2}+ \frac{1}{n}$ and $n$ is odd, $(\ref{eq:z_cond_coord_anti_coord})$ admits the two solutions  $z^*=\frac{n-1}{2n}$ and $z^{**} =\frac{n+1}{2n}$ (Fig. \ref{fig:eq_not_cohesive}), which are also discontinuous points.

The fractions associated to $z^*$ are $z^*_c = F_c^-(\frac{n}{n-1}\frac{n-1}{2n}) = F_c^-(\frac{1}{2}) = 0 $ and $z^*_a= z^* /(1-\alpha)$. Also in this case, the triple $(z^*, 0,z^* /(1-\alpha))$ satisfies (\ref{eq:nec_suff_cond}), and the same holds for the triple $(z^{**}, 1,(z^{**}-\alpha)/(1-\alpha))$.

CASE 3: If $\alpha < \frac{1}{2}+ \frac{1}{n}$ and $n$ is even, the unique solution of $(\ref{eq:z_cond_coord_anti_coord})$ is $z^*=\frac{1}{2}$ (Fig. \ref{fig:eq_not_cohesive}). Notice that
	$$
	F_c^-\left(\frac{n}{n-1}\frac{1}{2}\right) = 1 \neq F_c\left(\frac{n}{n-1}(\frac{1}{2}-\frac{1}{n})\right) = 0\,.
	$$
	Hence, $z^*$ does not induce a triple that solves (\ref{eq:nec_suff_cond}). By checking the discontinuous points, one can prove that, in this case, the mixed CAC game admits no NE.

\section{Concluding remarks}\label{conclusion}
We considered mixed ensembles of agents playing coordinating and anti-coordinating agents with possibly different activation thresholds. We found a necessary and sufficient condition for the existence of Nash equilibria in terms of the distribution of the thresholds in the ensemble and an explicit characterization of Nash equilibria when do exist. When only coordinating or anti-coordinating agents are present, we were capable to determine NE's cardinality.

Current work includes the extension of this analysis to the case when agents are confined to interact through a network. Preliminary results in this direction can be found in \cite{tesi} and \cite{laura}, where we determined graph-theoretical sufficient conditions for the existence of a
pure strategy Nash equilibria for mixed network CAC
games. Also, we are interested in the behavior of evolutionary dynamics associated to these games, such as the the best response dynamics.
\bibliography{bib} 
\appendix
\section{Proofs}
\begin{pf}(Proposition 1)
		The proof is based on Corollary 2.9  of \cite{monderer}. In our notation, it states that a strategic form game is a potential game if and only if for every two agents $i,j\in\V$, every action configuration $x_{-\{i,j\}}\in \A^{\V\setminus \{i,j\}}$ and every two possible actions of agent $i$, that is, $x_i, y_i\in \A$,  and of agent $j$, namely $x_j,y_j\in\A$, it holds
		$$
		\begin{aligned}
			u_i (B)&- u_i(A) + u_j(C) - u_j(B) \\&+ u_i(D)- u_i(C) + u_j(A)- u_j(D)= 0
		\end{aligned}
		$$
		where 
		$$
		\begin{aligned}
			A&= (x_i,x_j, x_{-\{i,j\}})\hspace{0.5cm}
			B= (y_i,x_j, x_{-\{i,j\}})\\
			C&= (y_i,y_j, x_{-\{i,j\}})\hspace{0.5cm}
			D= (x_i,y_j, x_{-\{i,j\}})\,.
		\end{aligned}
		$$
		We will show that the property does not hold if we pick a coordinating agent $i\in \V_c$ and an anti-coordinating agent $j\in \V_a$.
		Consider any configuration of the remaining agents $x_{-\{i,j\}}\in \A^{\V\setminus \{i,j\}}$ and let $x_i=x_j=1$ and $y_i =y_j=-1$. 
		Notice that
		$$
		\begin{aligned}
			u_i(A)&=u_{i}(1,1, x_{-\{i,j\}}) = 1+\sum_{k \neq i\,, j}x_k -d_i\\
			u_i(B)&=u_{i}(-1,1, x_{-\{i,j}\}) =-1-\sum_{k \neq i\,, j}x_k +d_i 
		\end{aligned}
		$$
		Therefore, the incentive of player $i$ in changing from action $1$ to $-1$ is given by
		$$
		u_i(B)-u_i(A)= -2-2\sum_{k \neq i\,, j}x_k +2d_i
		$$
		Let us now consider the utilities of the anti-coordinating agent $j$ for the actions $1$ and $-1$ when agent $i$ is playing $-1$:
		$$
		\begin{aligned}
			u_j(B)&=u_{j}(-1,1, x_{-\{i,j\}}) = 1-\sum_{k \neq i\,, j}x_k +d_i\\
			u_j(C)&=u_{j}(-1,-1, x_{-\{i,j}\}) =-1+\sum_{k \neq i\,, j}x_k -d_i 
		\end{aligned}
		$$
		We then find
		$$
		\begin{aligned}	
			u_j(C)-u_j(B)= -2+2\sum_{k \neq i\,, j}x_k -2d_i
		\end{aligned}	
		$$
		Similarly, we shall have
		$$
		\begin{aligned}
			u_i(D)-u_i(C)&= -2+2\sum_{k \neq i\,, j}x_k -2d_i\\
			u_j(A)-u_j(D)&= -2-2\sum_{k \neq i\,, j}x_k +2d_i
		\end{aligned}
		$$
		If we sum all the quantities, the result is
		$-8\neq 0\,.$
		 Therefore, the mixed CAC game is not a potential game when $\V_c \neq \emptyset$ and $\V_a \neq \emptyset$.
\end{pf}
\begin{pf}(Theorem \ref{prop:nec_suff_cond})
	\textit{Notation for proof.}
	Given an action configuration $x\in \X$, we denote with $\V^+(x):=\set{j\in \V \mid x_j = +1}$ the set of agents playing $1$ in $x$ and with $n^+(x):=\abs{\V^+(x)}$ its cardinality. We then define
	\begin{equation}\label{z_tilde}
		\tilde{z}(x) := 
		 \frac{n^+(x)}{n-1}\,,
	\end{equation}
	where the normalization factor is due to the fact that each agent interacts with $n-1$ opponents, that is, everybody but herself. Notice that, when an agent $i\in \V^+(x)$ is playing $1$, the fraction of opponents playing $1$ is given by
	\begin{flalign}\label{eq:complete_1}
		\frac{n^+_i(x)}{n-1} = \frac{n^+(x)-1}{n-1} = \tilde{z}(x)-\frac{1}{n-1}\,,
	\end{flalign}
	while if she plays $1$, that is, if $i\in\V^-(x):=\V\setminus\V^+(x)$,  we shall find
	\begin{equation}\label{eq:complete_2}
		\frac{n^+_i(x)}{n-1} 
		= \frac{n^+(x)}{n-1} =\tilde{z}(x) \,.
	\end{equation}  
Let us further denote with $\V^+_c(x):=\set{j \in \mathcal{V}_c\mid x_j =+1}$ the set of coordinating agents playing action $+1$ in $x$ and by $n^+_c(x):=\abs{\V^+_c(x)}$ its cardinality. Similarly, we shall define $\V^+_a(x):=\set{j \in \mathcal{V}_a\mid x_j =+1}$ and $n^+_a(x):= \abs{\V^+_c(x)}$. If $n_c >1$ and $n_a>1$, we can then define
	\begin{equation*}
		\tilde{z}_c(x) := \frac{n^+_c(x)}{n_c-1}\,,\hspace{1cm} \tilde{z}_a(x) := \frac{n^+_a(x)}{n_a-1}\,.
	\end{equation*}
	Observe that $\tilde{z}$ can be expressed as a combination of $\tilde{z}_c$ and $\tilde{z}_a$, that is,
	\begin{equation}\label{eq:cond_3}
		\tilde{z}(x) = \tilde{\alpha} \tilde{z}_c(x) + \left(\frac{n}{n-1}-\tilde{\alpha}\right) \tilde{z}_a(x)
	\end{equation}
where $\tilde{\alpha} = \frac{n_c -1}{n-1}$.
	Finally, we denote the re-normalized threshold CDF of the coordinating agents by
	$$\tilde{F}_c(z) := \frac{1}{n_c-1}\abs{\set{i \in \V_c \mid r_i \leq z}}\,,$$ while we denote the re-normalized threshold CCDF of the anti-coordinating agents by $$\tilde{G}_a(z) := \frac{1}{n_a-1}\abs{\set{i \in \V_a \mid r_i > z}}\,.$$ 
	\textit{Necesssary condition.}
	A substitution of $\eqref{eq:complete_1}$ and $\eqref{eq:complete_2}$ in (\ref{eq:br}) leads to the following condition on Nash equilibria
	\begin{equation}\label{ne_coord_anti_coord}
		x\in \mathcal{N} \hspace{0.3cm}\Leftrightarrow\hspace{0.3cm}
		\begin{cases}
			r_i \leq \tilde{z}(x) - \frac{1}{n-1}   \hspace{0.5cm} &i\in \V_c^+(x)\\
			r_i \geq \tilde{z}(x) &i\in \V_c^-(x)\\
			r_i \geq \tilde{z}(x) - \frac{1}{n-1}   \hspace{0.5cm} &i\in \V_a^+(x)\\
			r_i \leq \tilde{z}(x) &i\in \V_a^-(x)
		\end{cases}
	\end{equation}
	The first condition in (\ref{ne_coord_anti_coord}) allows to find an upper bound for the fraction of \textit{coordinating agents} playing action $+1$ in a Nash equilibrium  $x^*\in \X$, i.e. $\tilde{z}_c(x^*)$. Indeed, the number of coordinating agents choosing $+1$ in the Nash equilibrium is at most equal to the total number of coordinating agents having threshold less or equal than $\tilde{z}(x^*) - \frac{1}{n-1}$. In formulas, 
	\begin{flalign*}
		\tilde{z}_c(x^*)  &= \frac{n^+_c(x^*)}{n_c-1} \\&\leq \frac{1}{n_c-1}\abs{\set{i \in \V_c \mid r_i \leq \tilde{z}(x^*)-1/(n-1)}} \\ 
		&= \tilde{F}_c(\tilde{z}(x^*)-1/(n-1))\,.
	\end{flalign*}
	Similarly, the second condition in (\ref{ne_coord_anti_coord})
	can be used to find a lower bound for $\tilde{z}_c(x^*)$. Indeed, the number of coordinating agents choosing $-1$ in the Nash equilibrium is at most equal to the total number of coordinating agents having threshold greater or equal than $\tilde{z}(x^*)$, i.e., 
	\begin{flalign*}
		\frac{n_c}{n_c-1}-\tilde{z}_c(x^*) &= \frac{n^-_c(x^*)}{n_c-1} \\ &\leq  \frac{\abs{\set{i \in \V_c\mid r_i \geq \tilde{z}(x^*)}}}{n_c-1} \\
		&=\frac{n_c-\abs{\set{i \in \V_c \mid r_i < \tilde{z}(x^*)}}}{n_c-1}
		\\&=\frac{n_c}{n_c-1}-\tilde{F}^-_c(\tilde{z}_c(x^*))\,.
	\end{flalign*}
	If we combine the two, we shall find 
	\begin{equation}\label{cond_coord_diseq}
		\tilde{F}_c(\tilde{z}(x^*)-1/(n-1)) \geq \tilde{z}_c(x^*)  \geq \tilde{F}^-_c(\tilde{z}(x^*))
	\end{equation}
	Observe that, since $\tilde{F}_c$ is a non-decreasing function, we have that $\tilde{F}_c(\tilde{z}(x^*)-1/(n-1))\leq \tilde{F}^-_c(\tilde{z}(x^*))$. 
	The \textit{$1^{st}$ necessary condition} is then
	\begin{equation}\label{eq:cond_1}
				\tilde{F}_c(\tilde{z}(x^*)-1/(n-1)) =\tilde{z}_c(x^*) = 
		\tilde{F}^-_c(\tilde{z}(x^*)) 
	\end{equation}
Notice that the equality actually holds for any $z \in [\tilde{z}(x^*)-1/(n-1),\tilde{z}(x^*))$, that is, the function must be flat in such interval.
	
	We now aim to derive an upper bound and a lower bound for 
	$\tilde{z}_a(x^*)$. Notice that the number of anti-coordinating agents playing action $+1$ in $x^*$ is at most equal to the total number of agents having threshold greater or equal than $\tilde{z}(x^*) - \frac{1}{n-1}$ (third condition in \eqref{ne_coord_anti_coord}). In formulas,
	\begin{flalign*}
		\tilde{z}_a(x^*) &= \frac{n^+_a(x)}{n_a-1}  \\&\leq \frac{1}{n_a-1}\abs{\set{i \in \V_a \mid r_i \geq \tilde{z}(x^*)-1/(n-1)}}  
		\\&= \tilde{G}^-_a(\tilde{z}(x^*)-1/(n-1))\,.
	\end{flalign*}
	Similarly, according to the fourth condition of (\ref{ne_coord_anti_coord}), we have that the number of anti-coordinating agents playing action $-1$ in $x^*$ is at most equal to the total number of agents having threshold less or equal than $\tilde{z}(x^*)$, i.e.,
	\begin{flalign*}
		\frac{n_a}{n_a-1}-\tilde{z}_a(x^*) &= \frac{n^-_a(x)}{n_a-1}  \\ &\leq  \frac{1}{n_a-1}\abs{\set{i \in \V_a \mid r_i \leq \tilde{z}(x^*)}} \\&=\frac{n_a}{n_a-1}-\tilde{G}_a(\tilde{z}(x^*))
	\end{flalign*}
	We then find the $2^{nd}$ necessary condition
	\begin{equation}\label{eq:cond_2}
		\tilde{G}^-_a(\tilde{z}(x^*)-1/(n-1)) \geq \tilde{z}_a(x^*)  \geq \tilde{G}_a(\tilde{z}(x^*))\,.
	\end{equation}
	
	If we combine the necessary conditions in $\eqref{eq:cond_1}$ and $\eqref{eq:cond_2}$ and we recall the relation between $\tilde{z}_c$, $\tilde{z}_a$ and $\tilde{z}$ observed in $\eqref{eq:cond_3}$, we obtain the following \textit{necessary condition} for $x^*\in \mathcal{N}$ to be a Nash equilibrium:
	\begin{equation*}
		\begin{cases}
		\tilde{F}_c(\tilde{z}(x^*)-1/(n-1)) =\tilde{z}_c(x^*) = 
		\tilde{F}^-_c(\tilde{z}(x^*))   \\
	\tilde{G}^-_a(\tilde{z}(x^*)-1/(n-1)) \geq \tilde{z}_a(x^*)  \geq \tilde{G}_a(\tilde{z}(x^*)) \\
			\tilde{z}(x^*) = \tilde{\alpha} \tilde{z}_c(x^*) + \left(\frac{n}{n-1}-\tilde{\alpha}\right) \tilde{z}_a(x^*)
		\end{cases}
	\end{equation*}
	
		The system in Proposition \ref{prop:nec_suff_cond} is then obtained by applying the following substitutions 
		$$
	\begin{aligned}
	\tilde{z}(x^*)&= \frac{n}{n-1}z^*\,,\,\,	\tilde{z}_c(x^*)= \frac{n_c}{n_c-1}z_c^*\,,	\,\, \tilde{z}_a(x^*)= \frac{n_a}{n_a-1}z_a^*\,,\\
	&\tilde{F}_c(\cdot)=\frac{n_c}{n_c-1}F_c(\cdot)\,,\quad \tilde{G}_a(\cdot)=\frac{n_a}{n_a-1}G_a(\cdot)\,,	
\end{aligned}$$ 
plus the definition of $\alpha$ and $\tilde{\alpha}$ and by multiplying both members of the last equation by $\frac{n-1}{n}$.

	\textit{Sufficient condition.} 
	The conditions in (\ref{eq:nec_suff_cond}) guarantee that an action configuration built in the way presented in Remark 1 satisfies $z_c^*(x^*) =z^*_c$ and $ z_a^*(x^*)=z^*_a$. Then, it is possible to verify that such an action configuration, that in general is not unique, is a Nash equilibrium of the  mixed CAC game.
\end{pf}
\begin{pf}(Proposition \ref{pr:ex_card_coord})
	To prove existence it is sufficient to consider the potential function in \eqref{potential_function_het_coord}, as potential games always admit Nash equilibria. When no stubborn agents are present, that is, when $F(0)=0$ and $F(1)=1$,  we can find the same result by observing that $z^0= 0$ and $z^1=1$ are solutions of (\ref{eq:cond_coord}). 
	
	On the other hand, the fact that there is a one to one correspondence between the fractions $z^*\in\{0,1/n, \dots, 1\}$ satisfying $\eqref{eq:cond_coord}$ and the NE equilibria of the game is a direct implication of how Nash equilibria are constructed in (\ref{eq:build_eq_coord}). Since $|\{0,1/n, \dots, 1\}|=n+1$, the number of NE never exceeds $n+1$.
\end{pf}
\begin{pf}(Proposition \ref{pr:ex_z_anticoord})
		Let $z^*\in \set{0,\frac{1}{n},\dots, 1}$ satisfy \eqref{eq:z_cond_anti_coord}. We shall prove that it satisfies \eqref{eq:cond_anti_coord} as well. Notice that the right-hand inequality is satisfied by definition of $z^*$. On the other hand, the left-hand inequality is satisfied since $\text{Im}(G_a)= \set{0,\frac{1}{n},\dots, 1}$. Indeed,
	\begin{flalign*}
		z^*-\frac{1}{n} &<G_a\left(\frac{n}{n-1}(z^*- \frac{1}{n})\right) \Rightarrow \\z^* &\leq G_a\left(\frac{n}{n-1}(z^*- \frac{1}{n})\right) \overset{(1)}{\leq} G^-_a\left(\frac{n}{n-1}(z^*-\frac{1}{n})\right)
	\end{flalign*}
	where (1) holds true since the complementary cumulative distribution function is non-increasing. This concludes the proof. 
\end{pf}
\begin{pf}(Proposition \ref{prop:card_anticoord})
		We start by proving point $(1)$.
		Recall that $\abs{\mathcal{Z}}\geq 1$ as the game is potential. More precisely, we do know that there is at least one solution, which is $z^*$ satisfying \eqref{eq:z_cond_anti_coord}.  Such $z^*$ always exists and, according to Proposition \ref{pr:ex_z_anticoord}, it satisfies \eqref{eq:cond_anti_coord} as well. 
		
		In the following, we shall prove that any fraction $z < z^*$ or $z>z^*+\frac{1}{n}$ cannot be a solution. 		
		Indeed, when $z < z^*$, we shall have		$
		z < z^*\leq G_a(\frac{n}{n-1}z^*)\leq G_a(\frac{n}{n-1}z)\,,
		$ which contradicts condition \eqref{eq:cond_anti_coord}.
		Conversely, when $z>z^*+\frac{1}{n}$, it must hold true
		$z > z^*+\frac{1}{n}\geq G^-_a (z^*+\frac{1}{n})\geq  G^-_a (z+\frac{1}{n})$. Hence, $1\leq \abs{\mathcal{Z}}\leq 2$.
		
		We now give some insights of when there can be a second solution. Notice that the only remaining candidate is  $z^{**}:= z^* + {1}/{n}$.
	The right-hand inequality in (\ref{eq:cond_anti_coord}) is always satisfied as 
	$$z^{**}=z^*+ \frac{1}{n}  \overset{(1)}{\geq} G_a\left(\frac{n}{n-1}z^*\right)\overset{(2)}{\geq} G_a\left(\frac{n}{n-1}(z^{**})\right)\,, $$
	where $(1)$ holds by definition of $z^*$ and $(2)$ as $G_a$ is non-increasing. Then, $z^{**}$ is a solution of $\eqref{eq:cond_anti_coord}$ if and only if the left-hand inequality is satisfied, that is,
	\begin{flalign*}
		z^{**} \leq G_a^- \left(\frac{n}{n-1}(z^{**}-\frac{1}{n})\right) \Leftrightarrow 
		z^* +\frac{1}{n}\leq G^-_a \left(\frac{n}{n-1}z^*\right)\,.
	\end{flalign*}
Notice that a necessary condition for it to be satisfied is 
	\begin{flalign*} 
		G^-_a \left(\frac{n}{n-1}z^*\right)  -  G_a \left(\frac{n}{n-1}z^*\right) \geq \frac{1}{n} 
	\end{flalign*}
	which means that there must be a jump discontinuity point in $z^*$.
	If the jump is sufficiently big, then $\abs{\mathcal{Z}}=2$. Otherwise, $|\mathcal{Z}|=1$.

	We conclude by observing that, given $z$ satisfying (\ref{eq:cond_anti_coord}) and $\zeta_1$ and $\zeta_2$ defined as in (\ref{z1_z2}), the quantity $\binom{n(\zeta_2-\zeta_1)}{n(z-\zeta_1)}$ counts all the possible ways of setting the actions of the agents having thresholds $r_i \in [\frac{n}{n-1}(z^* - 
	\frac{1}{n}), \frac{n}{n-1}\,z^* ]$ in such a way that the condition $z(x)=z$ is met (see (\ref{eq:build_eq_anti_coord})). 
\end{pf}

\end{document}